# Analysis of Topology Based Routing Protocols for Vehicular Ad-Hoc Network (VANET)

**A B M Moniruzzaman**
Department of Computer Science and Engineering,
Daffodil International University, Dhaka, Bangladesh
abm.mzkhan@gmail.com

**Md. Sadekur Rahmann**
Department of Computer Science and Engineering,
Daffodil International University, Dhaka, Bangladesh
sadekur738@gmail.com

## ABSTRACT
Now-a-days vehicles are one of the most important parts of our life. We need them to cross distances in our everyday life. In this paper we discuss Vehicular Ad-Hoc Network (VANET) technology that can ensure the maintenance of traffic rules and regulation. By applying this technology we can save life, save time, corruption, vehicle security, avoid collision and so on. Vehicular Ad Hoc Network (VANET) is a part of Mobile Ad Hoc Network (MANET). Every node or vehicle can move freely and they will communicate each other by wireless technology in coverage. The main goal of this research is to study the existing routing protocols for ad-hoc network system and compared between AODV (Reactive) and DSDV (Proactive). We have studied different types of routing protocols such as topology based, position based, cluster based, geo-cast based and broadcast based. We have simulated and compared AODV (Reactive) and DSDV (Proactive) to find out their efficiency and detect their flaws.

## 1. Introduction
It's the nature of human to want to the break rules. But rules are necessary to ensure safety and leading a better life. And in this era we live in technology is a way to ensure the maintenance of the rules. Now-a-days vehicles are one of the most important parts of our life. We need them to cross distances in our everyday life. But drivers cause accidents and traffic jams by over-speeding and trying to takeover other cars. And that causes harm to the traffic system. In this paper we will discuss such a technology that can ensure the maintenance of traffic rules and regulation. By applying this technology we can save life, save time, corruption, vehicle security, avoid collision and so on. Everybody knows technology make a job more efficient. If we include technology in traffic system definitely it will become more efficient and stable. We can get benefit from VANET technology to ensure efficient traffic system and traffic safety. Without technology we can see lots of problems in traffic system. Vehicular Ad Hoc Network (VANET) is a part of Mobile Ad Hoc Network (MANET). Every node or vehicle can move freely and they will communicate each other by wireless technology in coverage. The communication may be node-to-node (N2N), node to multi node (N2MN), Node to Road Side Unit (N2RSU) and road side unit to node (RSU2N).

## 2. VANET
A Vehicular Ad-Hoc Network or VANET is a technology that uses moving cars as nodes in a network to create a mobile network. VANET turns every participating car into a wireless router or node, allowing cars approximately 100 to 300 meters of each other to connect and, in turn, create a network with a wide range. As cars fall out of the signal range and drop out of the network, other cars can join in, connecting vehicles to one another so that a mobile Internet is created [1].

VANET has some unique characteristics which make it different from MANET as well as challenging for designing VANET applications. Some of the main some unique characteristics are - high dynamic topology, Frequent disconnected network, Mobility modeling, Battery power and storage capacity, Communication environment, Interaction with onboard sensors [38].

VANET will provide the user with various application and features. Some of the main applications are- data transfer, warning, path prediction, traffic control, violence control, speed control, traffic density control, emergency contact [17].

## 3. Protocols

The characteristic of highly dynamic topology makes the design of efficient routing protocols for VANET is challenging [36]. Overall classification of VANET routing protocols has been shown in the figure-3.1.

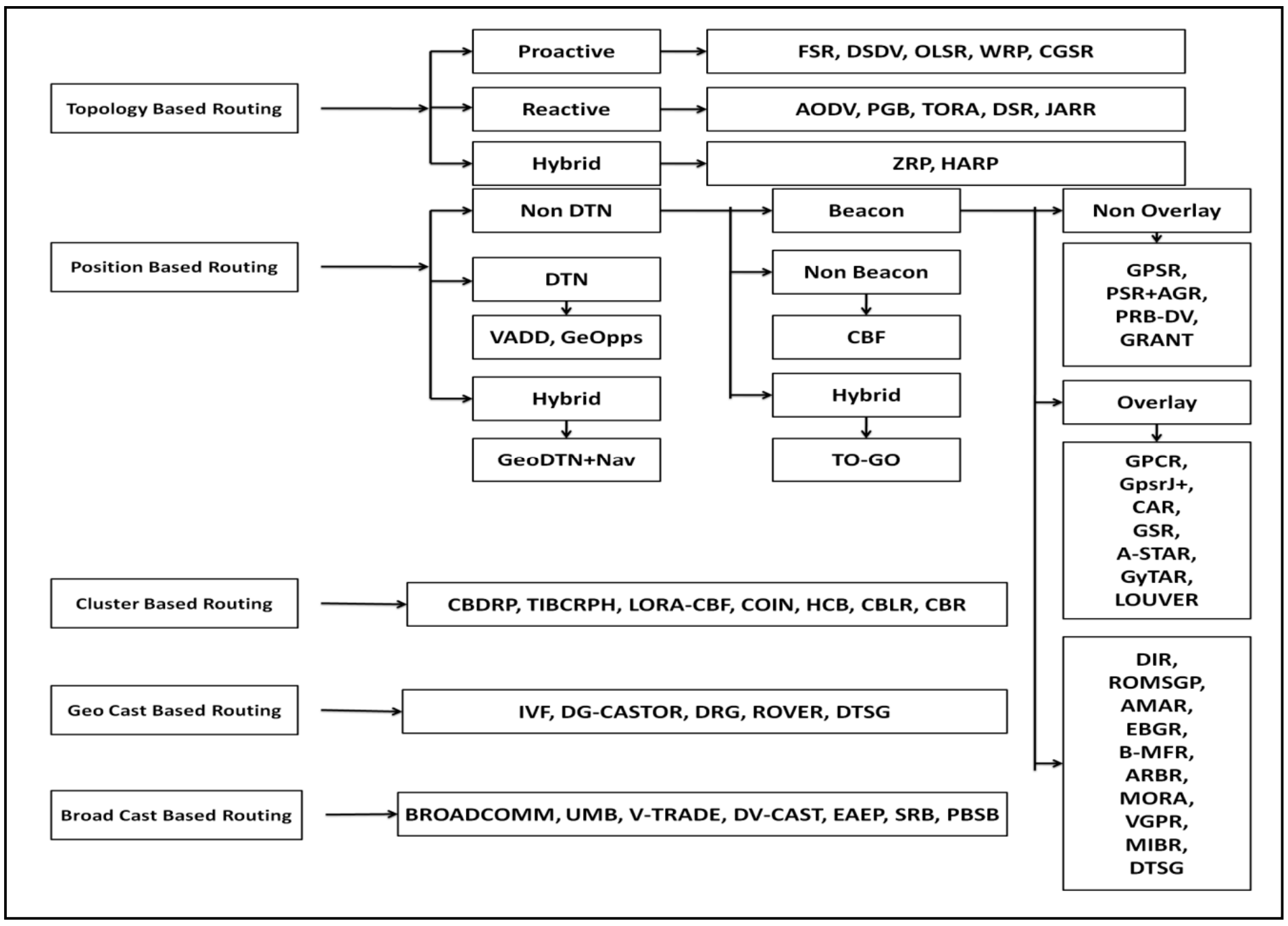

Fig 3.1: Types of routing protocol.

This routing protocol use link information that exists in the network to perform packet forwarding. There is three type of topology based routing - 1) Reactive. 2) Proactive, 3) Hybrid. Reactive routing protocol is called on demand routing because it starts route discovery when a node needs to communicate with another node thus it reduces network traffic.

*AODV: Ad Hoc on Demand Distance Vector*
In AODV [1] the network remains still until a connection is needed. When route is needed the source node broadcasts a request message. When the destination receives the request, it sends back a reply message through a temporary path to the source node. The source then begins connection using the route that has the least number of hops. The unused entries of the routing tables are erased after a time. If a link fails, then a routing error message is passed back to the source, and the process starts again. Each request for route has a different sequence number [2]. Nodes use this sequence number to avoid repeating a route requests that they have already passed on. Every route requests has a "time to live" number that limits the time that they can be retransmitted. If a route request fails, another request may not be sent instantly, it may wait until twice as much time has passed since the timeout of the previous request.

*Goals of AODV routing:* 1) Quick adaptation under dynamic link conditions. 2) Lower transmission latency. 3) Consume less network bandwidth (less broadcast). 4) Loop-free property. 5) Scalable to large network [2].

*An example of AODV routing*
In figure-3.2.there are five nodes, all of them are connected by bidirectional arrow. The starting node is S, from which position route will start.

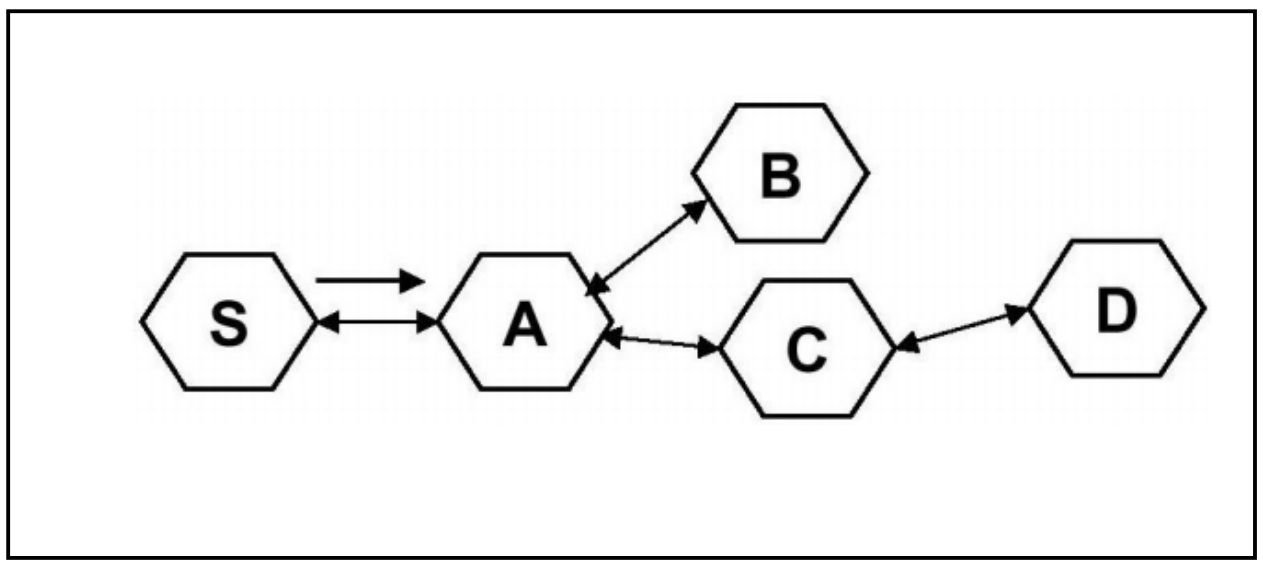

**Fig 3.2:** AODV routing [source: http://www.cs.stonybrook.edu/

*Proactive:* Proactive routing protocols are mostly based on shortest path algorithms. They keep information of all connected nodes in form of tables because these protocols are table based. Furthermore, these tables are also shared with their neighbors. Whenever any change occurs in network topology, every node updates its routing table.

Routing procedure of AODV step by step of figure-3.2.shown in below-

1. Node S needs a route to D
2. Create a route request (RREQ)
   - Enters D's IP address, sequence number, S's IP address, sequence number
   - Broadcasts RREQ to neighbors
3. Node A receives RREQ
   - Makes reverse route entry for S
     - o Dest = S, nexthop = S, hopcount = 0
   - It has no route to D, so it broadcasts RREQ
4. Node C receives RREQ
   - Makes reverse route entry for S
     - o Dest = S, nexthop = A, hopcount = 1
   - It has route to D &&seq# for route D >seq# in RREQ
     - o Creates a route reply (RREP)
   - Enters D's IP address, sequence number, S's IP address, hopcount
     - o Unicasts RREP to A
5. Node A receives RREP
   - Unicasts RREP to S
   - Makes forward route entry to D
     - o Dest = D, nexthop = C hopcount = 2
6. Node S receives RREP
   - Makes forward route entry to D
     - o Dest = D, nexthop = A hopcount = 3
   - Sends data packets on route to D

*DSDV: Destination-Sequenced Distance-Vector:* DSDV [1] is a table-driven routing scheme for ad hoc mobile networks based on the Bellman-Ford algorithm. Each entry in the routing table has a sequence number [2], if a link is present then the sequence numbers are even; otherwise an odd number is used.

The sequence number is generated by the destination, and the emitter must send out the next update with this number. Routing information is distributed between nodes through sending full dumps infrequently and smaller incremental updates more frequently. This routing adds two things to distance-vector routing: 1) Sequence number that helps avoiding loops. 2) Damping that holds the advertisements for changes of short duration.

*Update technique:* 1) Each node periodically transmits the updates, which includes its own sequences number, routing table updates.2)It also send the routing table updates for important link changes. 3)When two routes to a destination is received from two different neighbors it chooses the one with greatest destination sequence number but if sequence numbers are equal, choose the smaller metric (hop count).[2]

*Routing update is done in two ways- 1) Full dump:* that entire routing table is sent to the neighbor. 2) Transmits relatively infrequently when no movement of node occurs.3) Appropriate when the network change is more frequent.[2[[12] and 2) *Incremental:* 1)The entries that require changes are sent.2)Transmitted more frequently.3)Appropriate when the network is relatively stable.[2[[12]

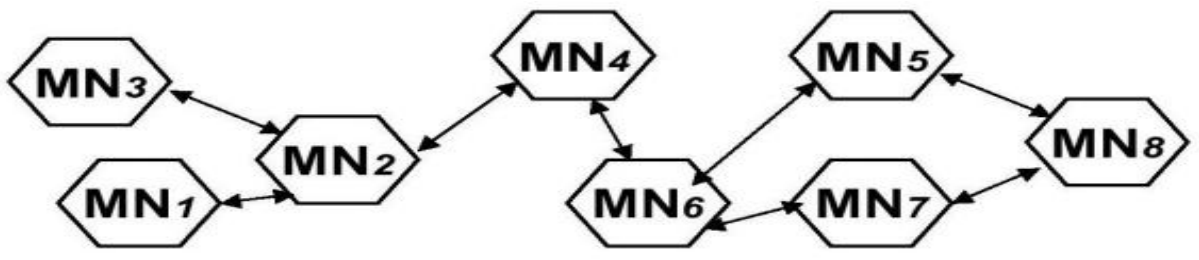

| Dest | Nexthop | Metric | DestSequence | InstallTime |
|---|---|---|---|---|
| MN1 | MN2 | 2 | 406 | |
| MN2 | MN2 | 1 | 128 | |
| MN3 | MN2 | 2 | 564 | |
| MN4 | MN4 | 0 | 710 | |
| MN5 | MN6 | 2 | 392 | |
| MN6 | MN6 | 1 | 076 | |
| MN7 | MN6 | 2 | 128 | |
| MN8 | MN6 | 3 | 050 | |

**Fig 3.3:** DSDV routing

## 4. Implementation

In this chapter we will discuss about implementation details. The features described with necessary figures. This chapter is organized as follows:

**4.1. Simulator and analyzer**

Actually a VANET simulator has two parts. One is network component and it is capable of simulating the behavior of a wireless network. Second is vehicular traffic component and this is capable to provide an accurate mobility model for the nodes of a VANET. Depending on the simulation the simulator can contain others components. To describe them, we refer to the NS2[27],[28],[29],[30] or NS3[32],Vanet MobiSim[31], Qualnet[23],[24] Sumo Simulator. We have worked on NS2. To analyze traffic we refer NS2-VisualTraceAnalyzer, Xgraph and trace analyzer. In our work we have used NS2-VisualTraceAnalyzer and Xgraph. We have use Comodo edit 7 to write simulation code. For coding we have used otcl and c++ language.

**4.2 Simulation environment**

The simulation environment created on Fedora 20. To make the simulation model we used NS2 simulator [27][28][29][30]. The NS2 simulator instruction has been use to define the topology structure of the network. For node configuration we used otcl [19][20][21][22][25][26][27] language instructions.

**4.3 Traffic model**

The source and destination are spread randomly over the network. We have used TCP with source node and TCPSink with destination node. We have attach FTP data source with TCP [11]. TCP packet size 512B and TcpSink packet size 210B. The maximum data source packet is 2048. Link bandwidth is 10Mbps. According to the number of source-destination pairs can be varied to the packet-sending rate in each pair[13].

**4.4 Mobility model**

The model uses the random way point[33] model in a rectangular field. The field configurations are 3000 m × 1600 m field with 100 nodes. Here, each packet starts its journey from a random location to a random destination. Nodes are moving at the speed of 0-84m/s. Sources and destinations are changing with respect of time. Simulations are run for 600 simulated seconds. Identical mobility and traffic scenarios are used across protocols to gather fair results.

**4.5 Communication models**

Communication models highlight the information flows between two vehicles and other moving object[10]. VANET applications are affected by[14, 15, 16] wireless networking aspects such as throughput, jitter, congestion window, bandwidth, transmission delay, packet loss or network access scheme. However, accurate network simulation introduces additional complexity and makes several large-scale VANET applications unsuitable for simulation.

**4.6 Traffic flow diagram**

Traffic flow direction and data source for our simulation.

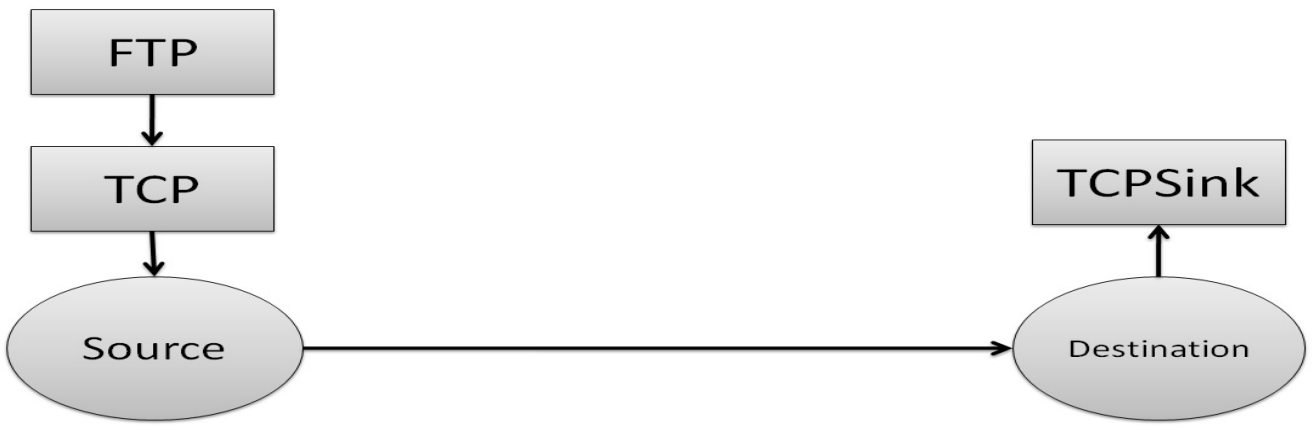

**Figure 4.1:** Traffic flow diagram

There are TCP with source node and TCP Sink with destination node. There are attach FTP data source with TCP.

**4.7 Design interface & Scenario**

Figure-4.2 shows 3000 m × 1600 m field with 100 nodes. Here, each packet starts its journey from a random location to a random destination. Nodes are moving at the speed of 0-84m/s. Sources and destinations are changing with respect of time. Simulations are run for 600 simulated seconds.

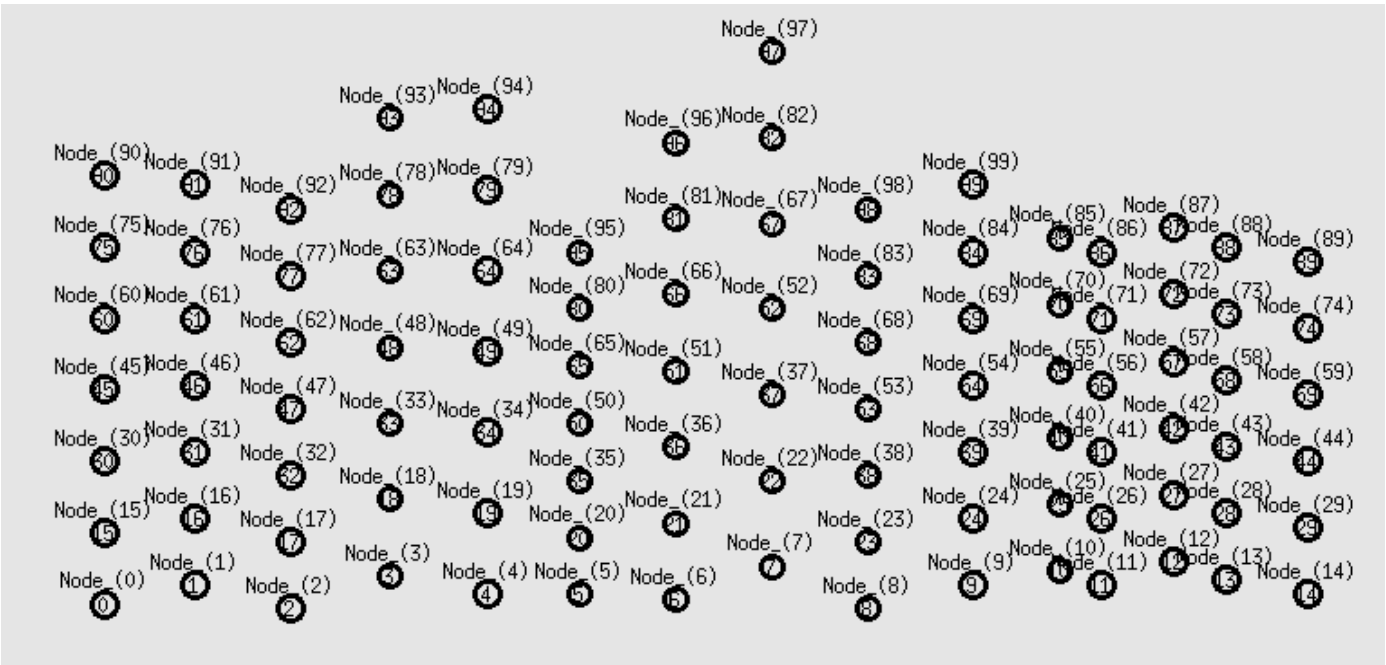

**Figure 4.2:** Simulation aria scenario

**4.8 Trace file Sample**

When we compile TCL code of protocols it produce some files i.e. .nam, .trce etc. In figure-4.3 we just show a screenshot of trace file.

```
M 0.00000 2 (550.00, 290.00, 0.00), (550.00, 290.00), 0.00
M 0.00000 17 (550.00, 430.00, 0.00), (550.00, 430.00), 0.00
M 0.00000 32 (550.00, 570.00, 0.00), (550.00, 570.00), 0.00
M 0.00000 47 (550.00, 710.00, 0.00), (550.00, 710.00), 0.00
M 0.00000 62 (550.00, 850.00, 0.00), (550.00, 850.00), 0.00
M 0.00000 77 (550.00, 990.00, 0.00), (550.00, 990.00), 0.00
M 0.00000 92 (550.00, 1130.00, 0.00), (550.00, 1130.00), 0.00
M 0.00000 3 (755.00, 360.00, 0.00), (755.00, 360.00), 0.00
M 0.00000 18 (755.00, 520.00, 0.00), (755.00, 520.00), 0.00
M 0.00000 33 (755.00, 680.00, 0.00), (755.00, 680.00), 0.00
M 0.00000 48 (755.00, 840.00, 0.00), (755.00, 840.00), 0.00
M 0.00000 63 (755.00, 1000.00, 0.00), (755.00, 1000.00), 0.00
M 0.00000 78 (755.00, 1160.00, 0.00), (755.00, 1160.00), 0.00
M 0.00000 93 (755.00, 1320.00, 0.00), (755.00, 1320.00), 0.00
M 0.00000 4 (960.00, 320.00, 0.00), (960.00, 320.00), 0.00
M 0.00000 19 (960.00, 490.00, 0.00), (960.00, 490.00), 0.00
M 0.00000 34 (960.00, 660.00, 0.00), (960.00, 660.00), 0.00
M 0.00000 49 (960.00, 830.00, 0.00), (960.00, 830.00), 0.00
M 0.00000 64 (960.00, 1000.00, 0.00), (960.00, 1000.00), 0.00
M 0.00000 79 (960.00, 1170.00, 0.00), (960.00, 1170.00), 0.00
M 0.00000 94 (960.00, 1340.00, 0.00), (960.00, 1340.00), 0.00
M 0.00000 5 (1150.00, 320.00, 0.00), (1150.00, 320.00), 0.00
M 0.00000 20 (1150.00, 440.00, 0.00), (1150.00, 440.00), 0.00
M 0.00000 35 (1150.00, 560.00, 0.00), (1150.00, 560.00), 0.00
M 0.00000 50 (1150.00, 680.00, 0.00), (1150.00, 680.00), 0.00
M 0.00000 65 (1150.00, 800.00, 0.00), (1150.00, 800.00), 0.00
M 0.00000 80 (1150.00, 920.00, 0.00), (1150.00, 920.00), 0.00
```

**Figure 4.3:** Trace file sample

## 5. Experimental Setup and Result

In this chapter we will discuss about Experimental results. These Experimental setup and results described with necessary figures. For simulation, I have used same configuration for both topology. I have use 100 wireless node five TCP connection and traffic source is FTP. TCP packet size 512B and TCP Sink packet size 210B. The maximum data source packet is 2048. Link bandwidth is 10Mbps. I have analyzed Throughput, Jitter, Delay, Congestion window and bandwidth for destination window with graph. Now i am presenting the comparison between proactive (DSDV) (Proactive) and reactive (AODV) (Reactive) protocol in VANET technology.

### 5.1 Node in long distance

For Node 0(Source) and Node 15(Destination), here node 0 and node 15 are in same coverage and both is moving object and they start their journey at 10s node 0 moves to node 15 position at the speed of 75m/s and node 15 move 2648m from its position at the speed of 12.97m/s. When time is 12s node 0 reached to node 15 position and node 15 become 25.94 far from node 0. At the time of 29.73s node 15 become out of range from node 0 for first time and connection is drop. In reactive (AODV) connections reconnect by changing the routing but in proactive (DSDV) connection can't reconnect due to table update.

### 5.1.1 Throughput transferred

Throughput describes the loss rate as seen by the transport layer. It reflects the completeness and accuracy of the routing protocol. From these graphs it is clear that throughput de-crease with increase in mobility. As the packet drop at such a high load traffic is much high [34].

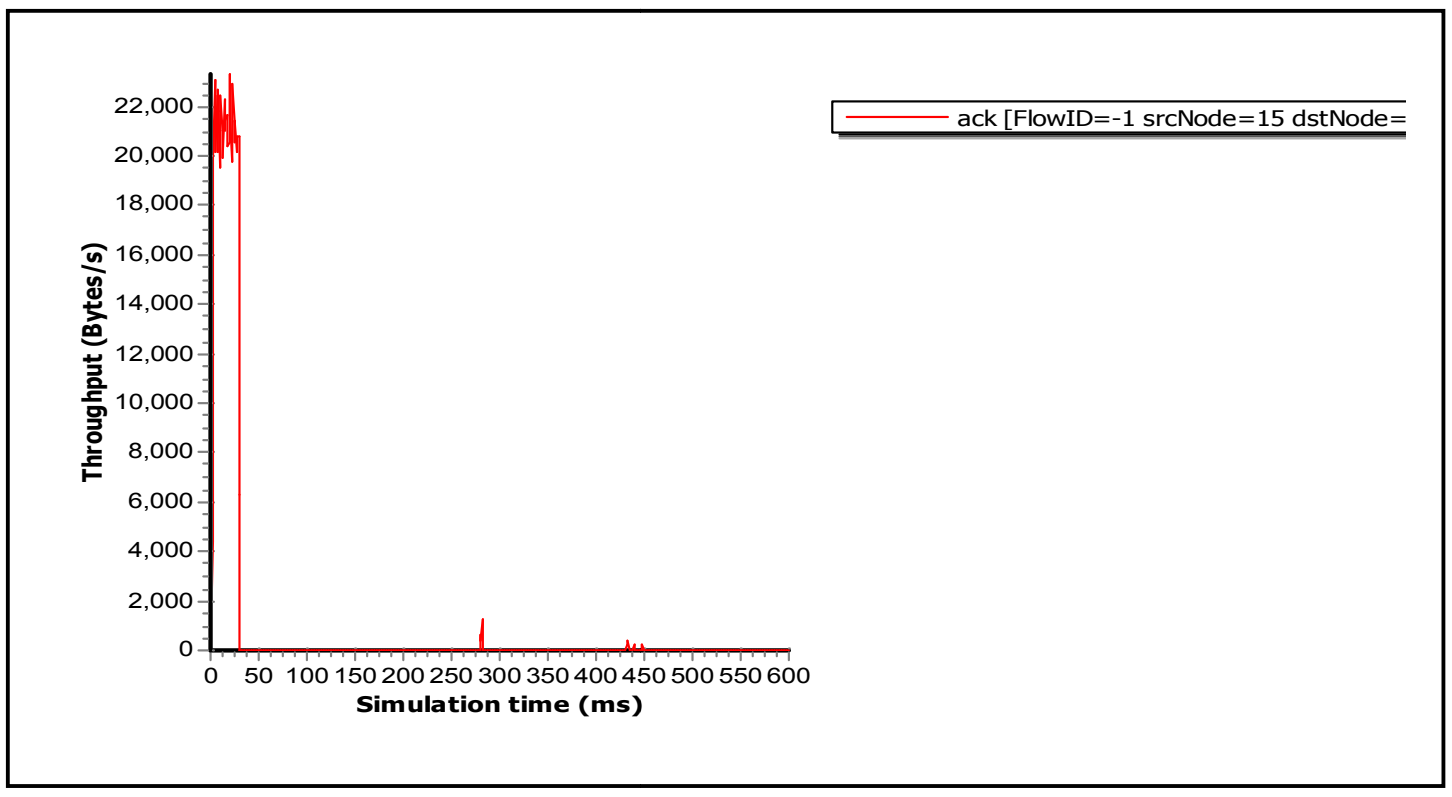

**Figure 5.1:** Throughput transfer from node 15 to node 0 in DSDV

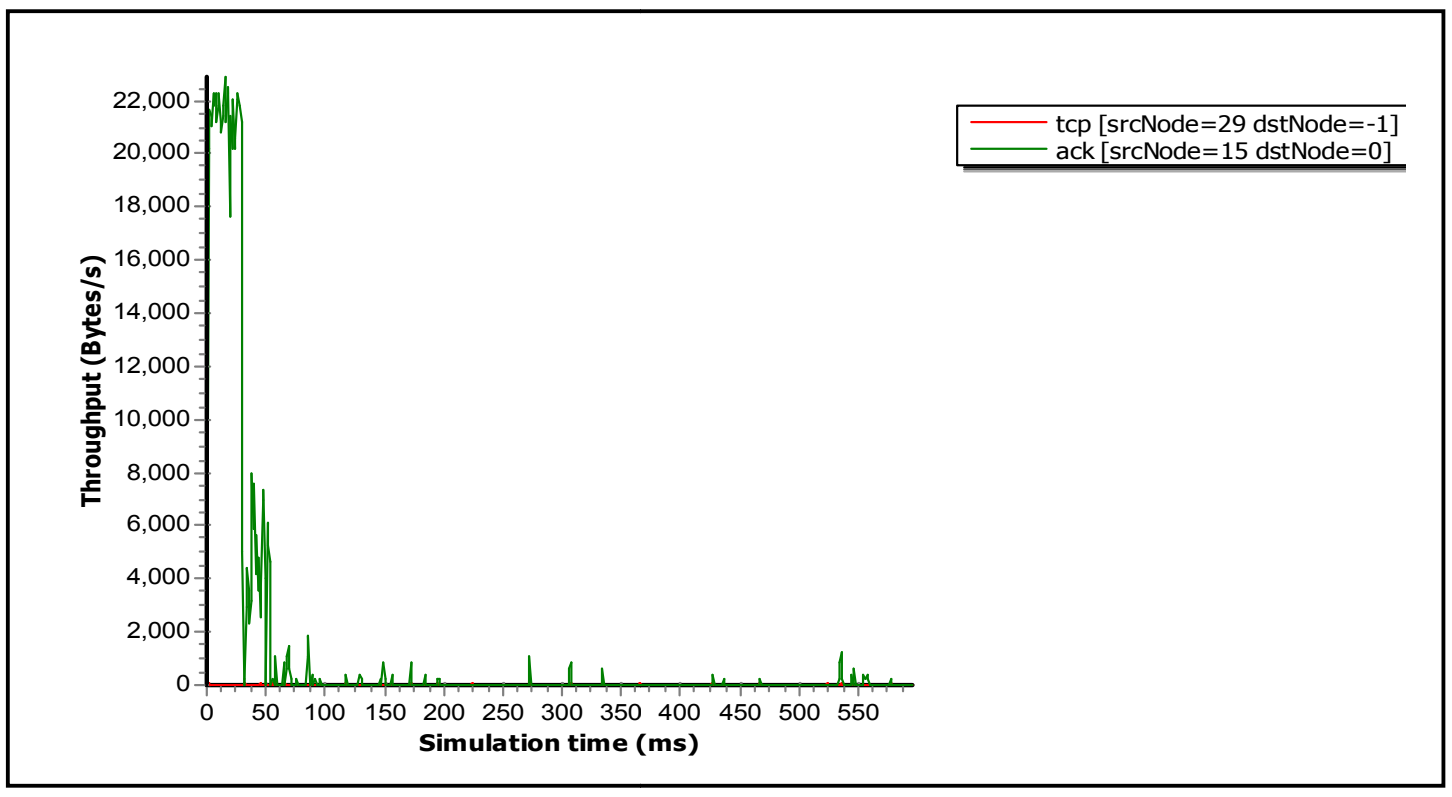

**Figure 5.2:** Throughput transfer from node 15 to node 0 in AODV

For figure 5.1&5.2 0s to 30s throughput curve was almost same for proactive (DSDV) and reactive (AODV). After 30s in reactive (AODV) throughput fluctuates for many times but in proactive (DSDV) throughput curve almost flat.

### 5.1.2 Jitter transferred

Jitter is the standard deviation of packet delay between all nodes and object.

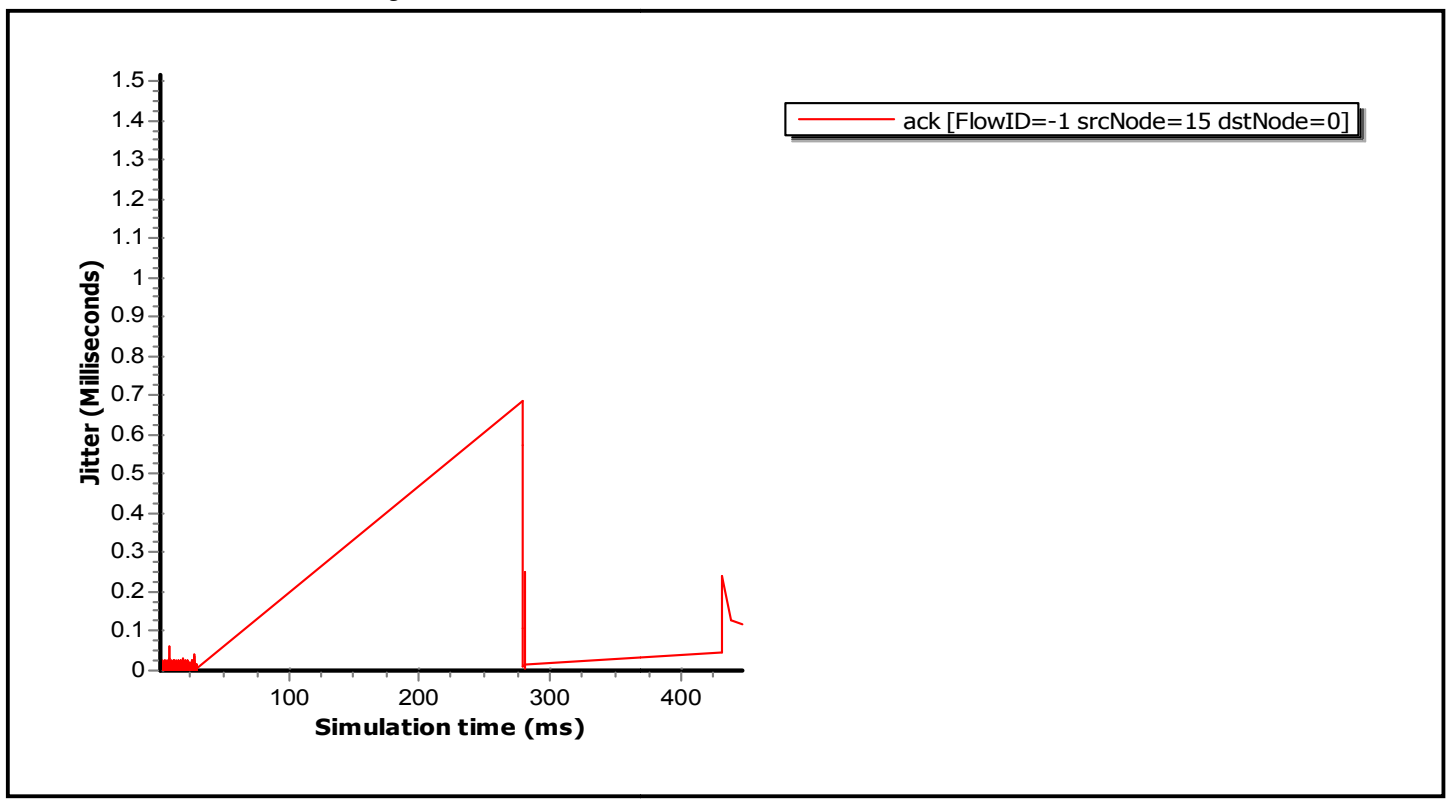

**Figure 5.3:** Jitter transfer from node 15 to node 0 in DSDV

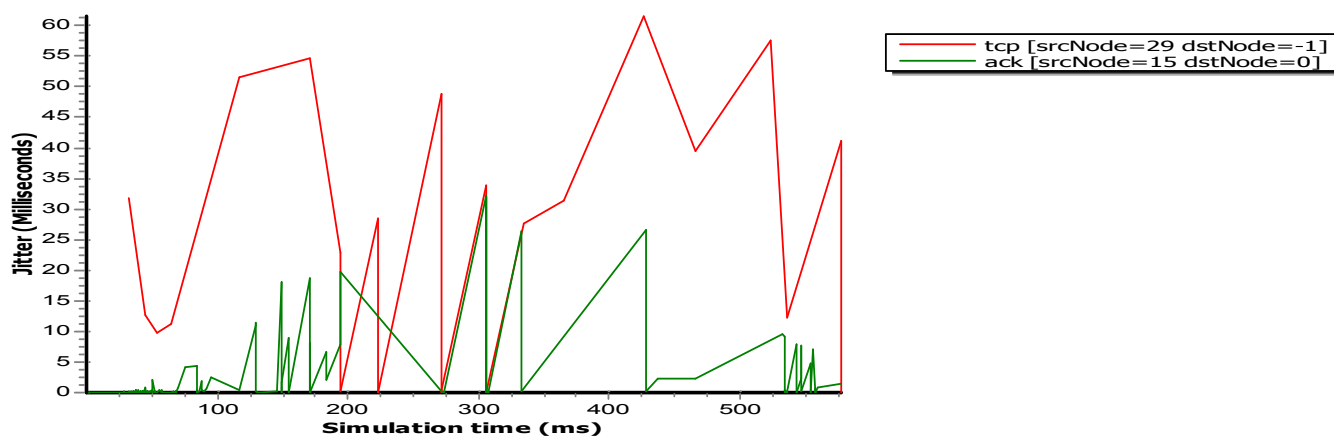

**Figure 5.4:** Jitter transfer from node 15 to node 0 in AODV

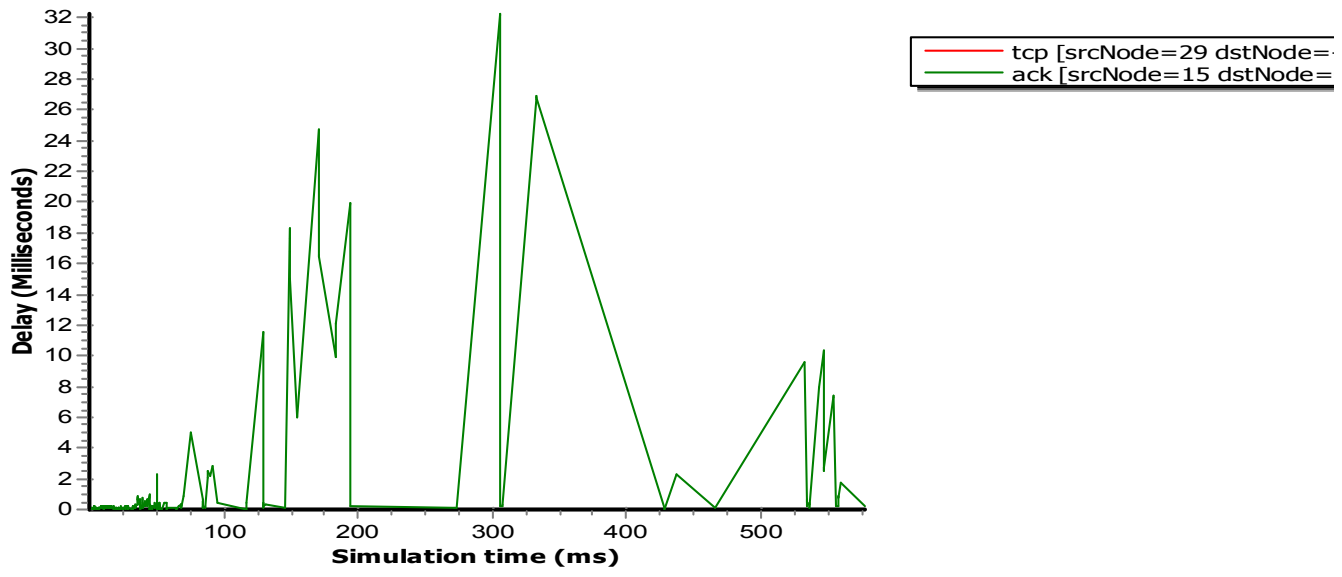

**Figure5.5:** Delay transfer from node 15 to node 0 in DSDV

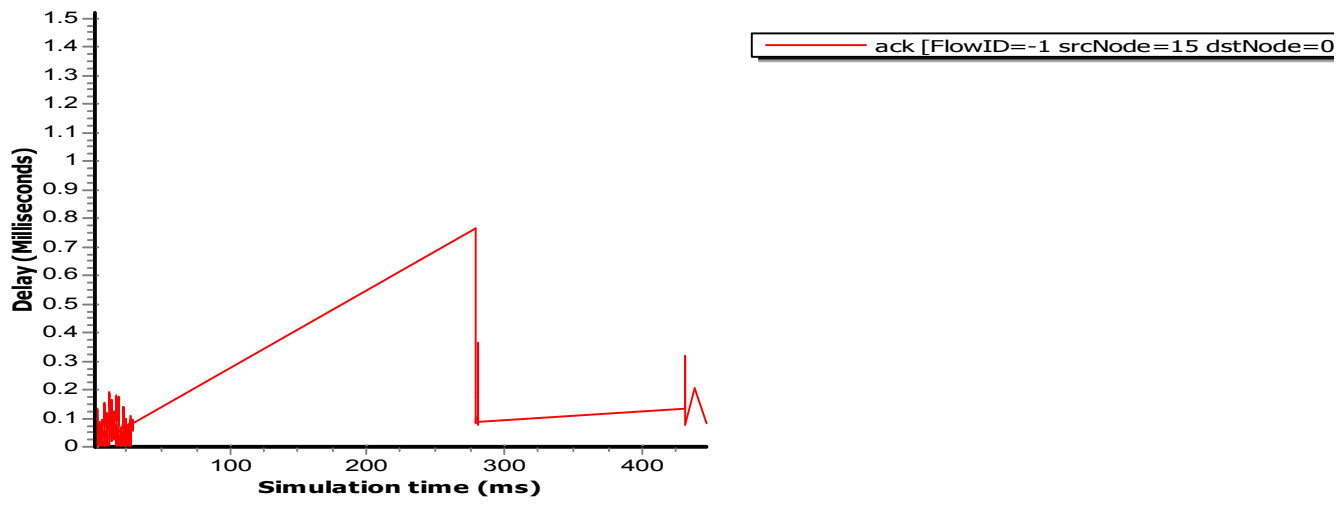

**Figure 5.6:** Delay transfer from node 15 to node 0 in AODV

For figure 5.3 and 5.4 jitter of proactive (DSDV) is less than reactive (AODV) and jitter fluctuates for several times in AODV. The maximum jitter is about 0.7 ms in proactive (DSDV) but the maximum jitter is about 35 ms in AODV. Overall the jitter of proactive (DSDV) is less than AODV.

### 5.1.3 Delay transferred

Delay is high for those nodes that are moving fast. AODV generate many ACK packets, so the ACK packet in AODV is indeed a great bottleneck [35]. As it was reactive protocol, so the delay little bit more.
For figure 5.5&5.6 delay of proactive (DSDV) is less than AODV. In proactive (DSDV) maximum delay is 0.8 ms but in reactive (AODV) maximum delay is about 32 ms. We know the delay of reactive is more proactive because reactive topology use a route discovery packet to find the destination as a result it takes few moment to reach the destination. On the other hand, in proactive source know the destination position before routing happen.

### 5.1.4 Congestion window

We can see in the figure at the beginning both curve go in same way. After that the curve of Proactive (DSDV) is almost flat and the curve of reactive (AODV) rises for several times. Overall the number of getting ACK in reactive (AODV) is higher then proactive(DSDV).

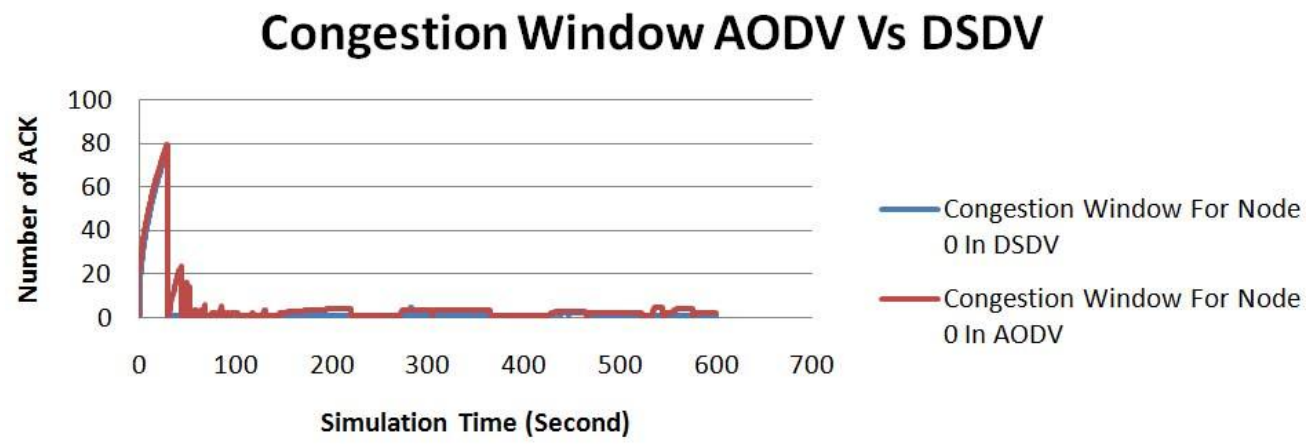

**Figure 5.7:** Congestion window reactive (AODV) & proactive (DSDV) for node 0

### 5.1.5 Bandwidth in destination Node

We can see in the figure the uses of bandwidth in proactive (DSDV) is less then reactive (AODV). This node was moving object as a result in proactive (DSDV) node left the neighbour nodes aria before update the routing table. For this reason it makes few data trajection.

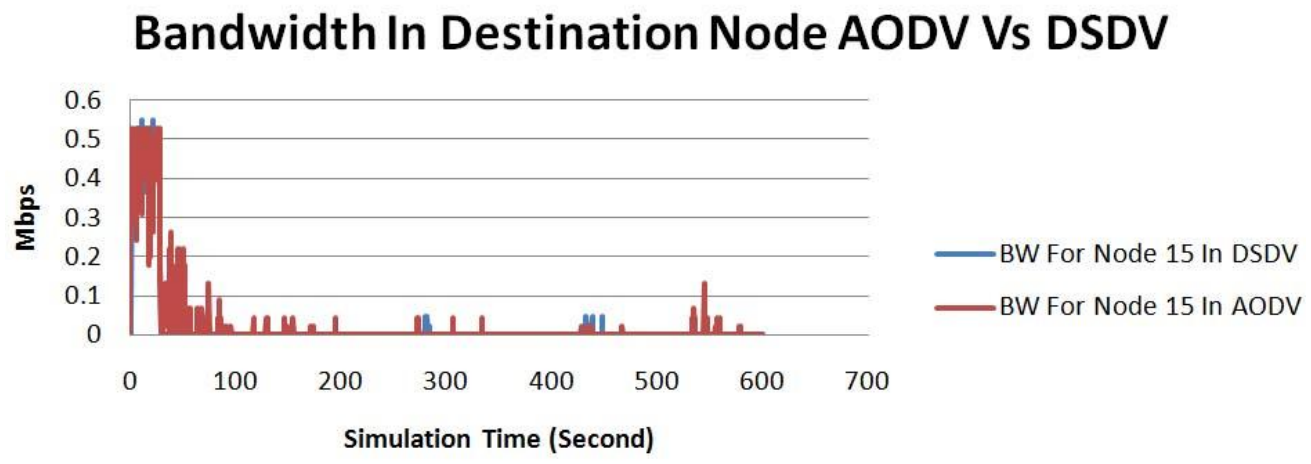

**Figure 5.8:** Bandwidth in destination node reactive (AODV) & proactive (DSDV) for node 15

### 5.2 Node in short distance

For Node 1(Source) and Node 25(Destination), here node 1 and node 25 are in 9 hop distance. Both are moving object and they start their journey at 10s node 1 moves to opposite direction from node 15 position at the speed of 12.66m/s and its go out of range at 24.36s. At the time of 153s node 28 and 5 comes to its range and connection reconnect.

In proactive (DSDV) at the time of 93.02s route happens according to the 1-28-25 direction but this packet has been lost because in the routing table of node 28 do not locate the right path for destination. At the time of 153s first successful route happens according to the 1-5-16-31-46-61-76-90-25 direction. In reactive (AODV) at the time of 153s first successful route happens according to the 1-28-5-16-32-33-48-63-77-76-90-25 direction.

### 5.2.1 Throughput transferred:

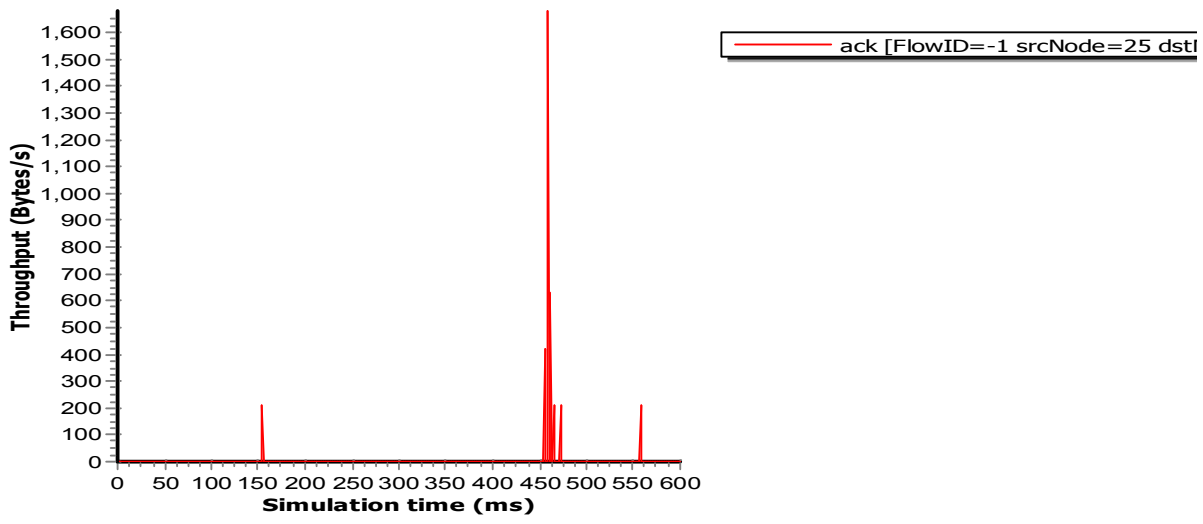

Figure 5.9: Throughput transfer from node 25 to node 1 in DSDV

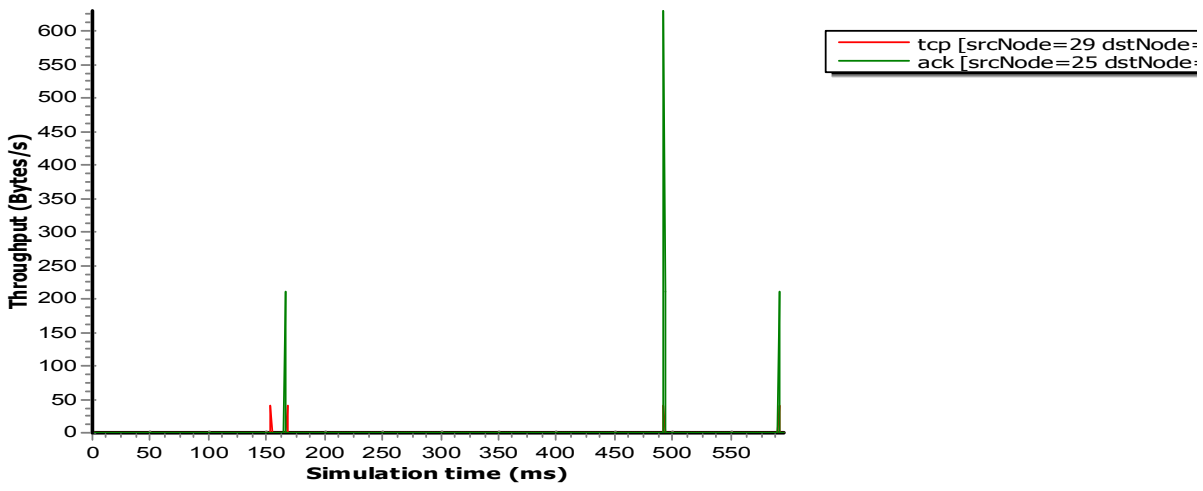

Figure 5.10: Throughput transfer from node 25 to node 1 in AODV

For figure 5.9&5.10 throughput starts at form 153s first time the transaction of throughput in proactive (DSDV) is higher than reactive (AODV) but in reactive (AODV) the throughput curve rises after few second. For this TCP connection the throughput high for proactive and it's happen after stop moving the object.

### 5.2.2 Jitter transferred

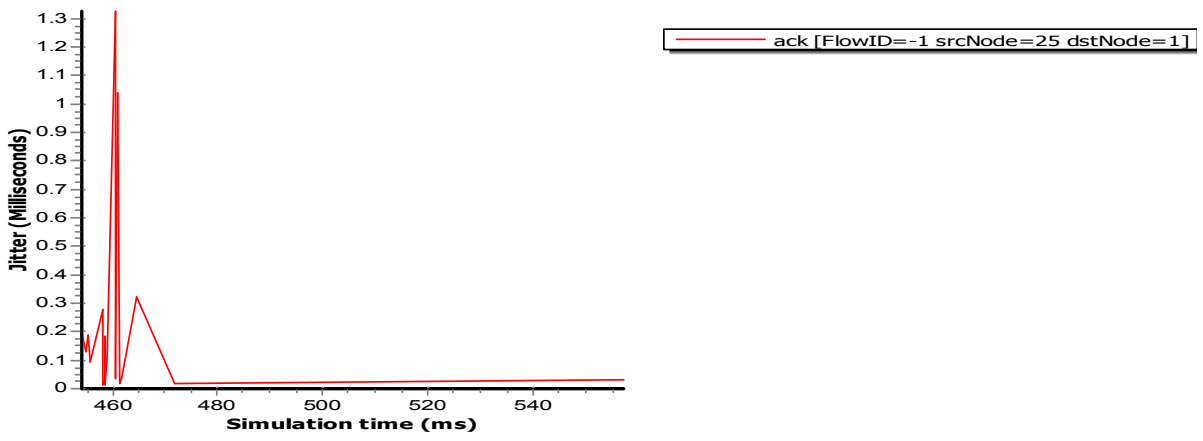

**Figure 5.11:** Jitter transfer from node 25 to node 1 in DSDV

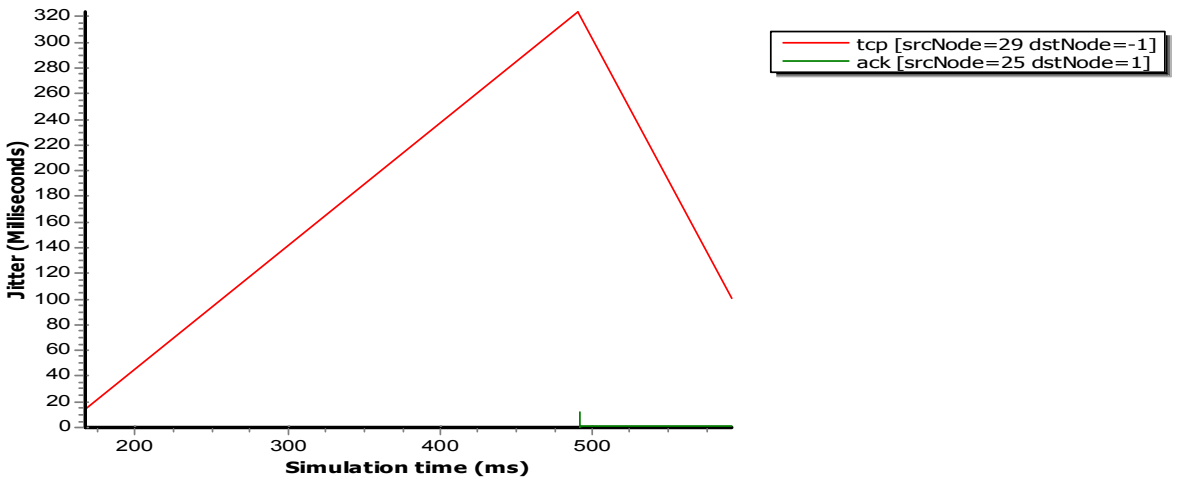

**Figure 5.12:** Jitter transfer from node 25 to node 1 in AODV

For figure 5.11&5.12jitter curve increase gradually in reactive (AODV). The minimum jitter of reactive (AODV) is higher than proactive (DSDV) jitter. Jitter represents the standard deviation of packet delay between nodes and object. Therefore, the standard deviation of this communication is high in reactive then proactive.

### 5.2.3 Delay transferred

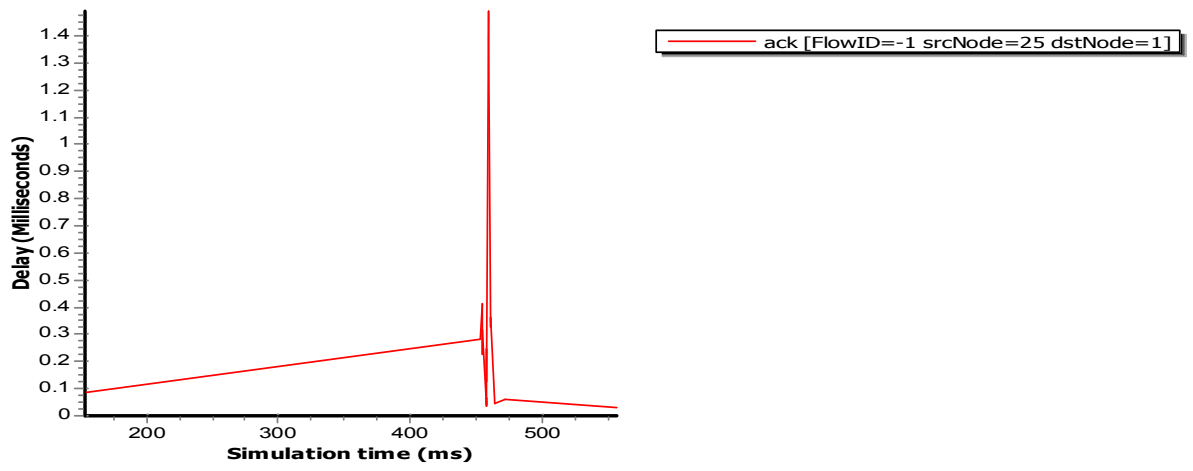

Figure 5.13: Delay transfer from node 25 to node 1 in DSDV

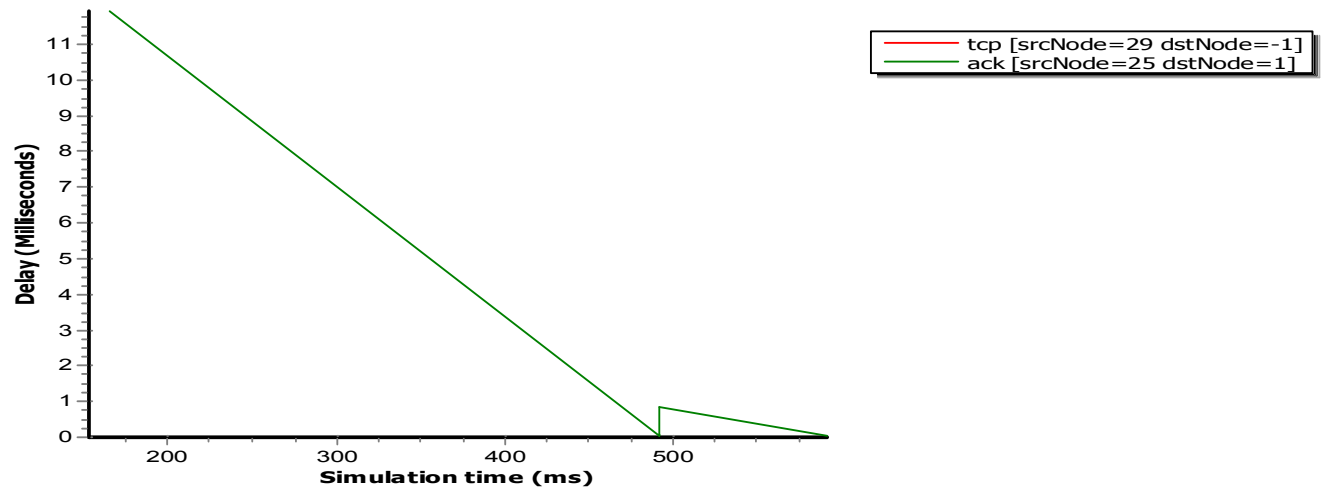

Figure 5.14: Delay transfer from node 25 to node 1 in AODV

For figure 5.13&5.14 the delay of reactive (AODV) is higher than proactive (DSDV). Delay curve have been decrease with time in reactive (AODV). In proactive delay is very low.

### 5.2.4 Congestion window

In the figure number of getting ACK curve increase after a time interval. Before stop the node the communication was stop because the node was moving so fast.

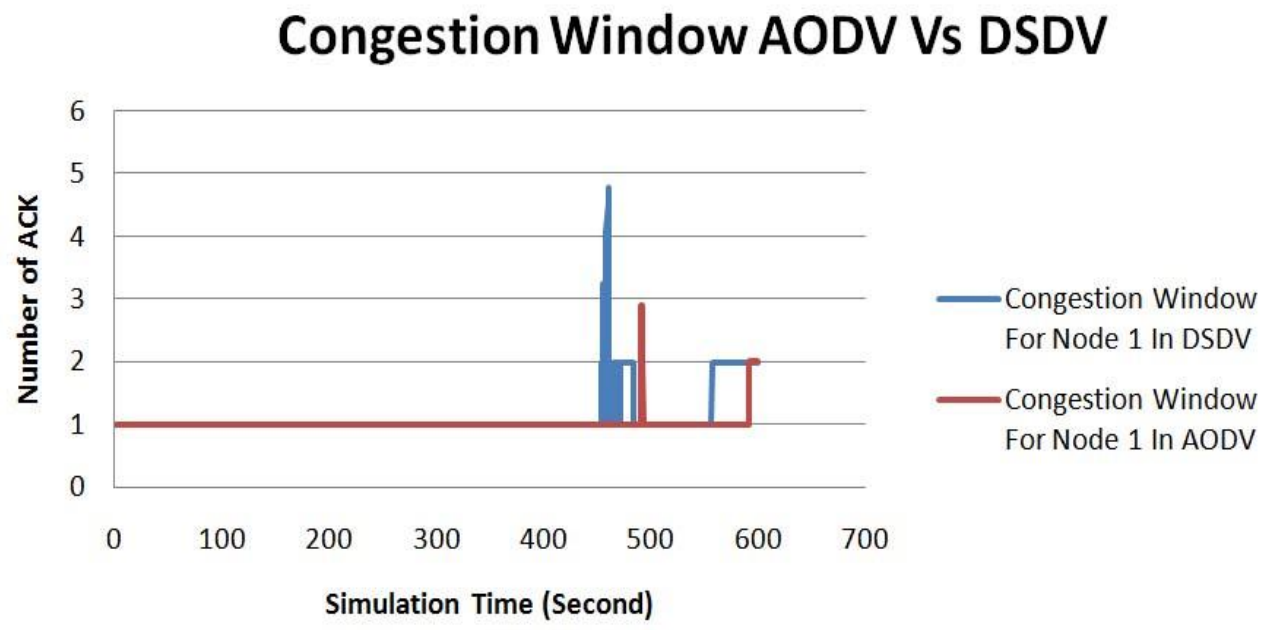

Figure 5.15 : Congestion window reactive (AODV) & proactive (DSDV) for node 1

After stop moving the node was 5 hop next from source node. As a result proactive (DSDV) build connection before reactive (AODV). Reactive connect the communication after few moment. It also represents the delay of reactive communication.

### 5.2.5 Bandwidth in destination Node

In the figure we can see the curve of reactive (AODV) rises fast then after about 100s proactive curve rises.

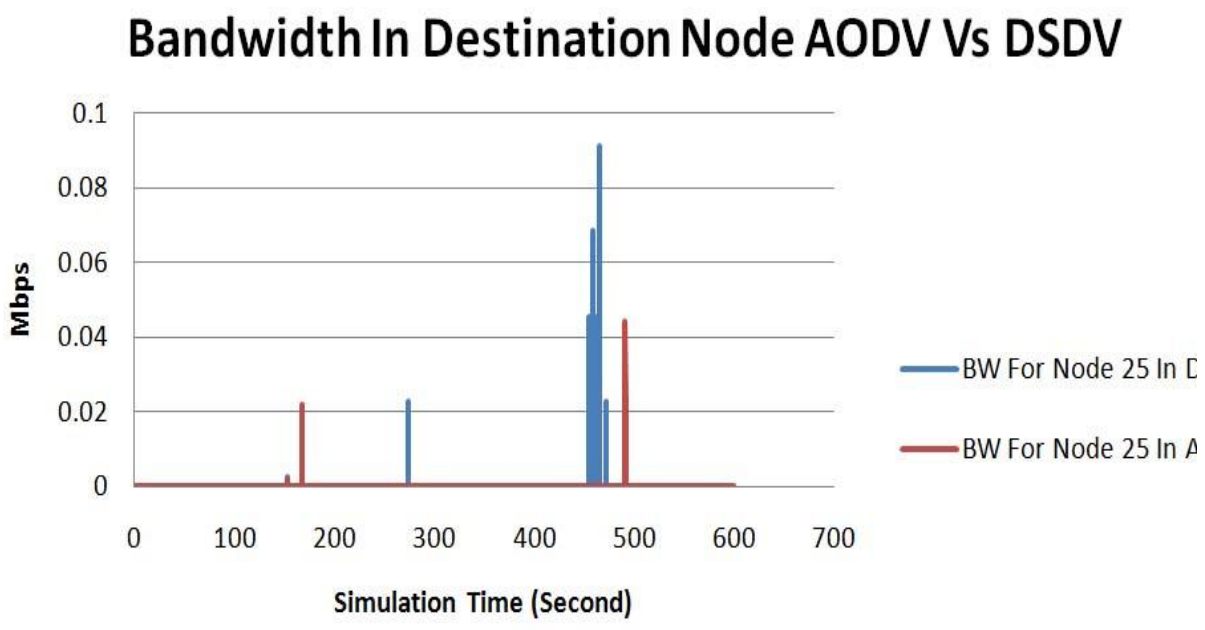

Figure 5.16 : Bandwidth in destination node reactive (AODV) & proactive (DSDV) for node 25.

However, at the end of simulation proactive used more bandwidth then reactive.

### 5.3 Observation

Some observations came from experimental results, these are as follows: 1) Reactive is more efficient then proactive for this technology. 2) Proactive is good for still object. 3) Delay of reactive procedure is more than proactive. 4) Reactive make unnecessary route. 5) Sometimes reactive don’t use the minimum path. 6) The data loss of reactive is more than proactive. 7) Reactive can’t route properly in long distance. 8) Proactive takes more time to update its table in long distance. 9) In short distance - proactive perform better then reactive. 10) In short distance the throughput & congestion window of proactive is higher than reactive but in long distance throughput & congestion window of reactive higher then proactive. 11) Both topologies don't work after a speed. Proactive or reactive is not suitable routing procedure on moving object.

## 7. Conclusion

In ad-hoc network community VANET has fascinated many researchers due to its distinctive nature. A huge amount of research has been made in various routing sectors of VANET; still there are many areas that need more research. The main goal of this research is to study the existing routing protocols for ad-hoc network system and compared between AODV (Reactive) and DSDV (Proactive). In our research we have simulated and compared AODV (Reactive) and DSDV (Proactive) to find out their efficiency and detect their flaws. In future we want to continue our research to learn more about the possibilities that lies within VANET.